\documentstyle[preprint,aps]{revtex}
\begin{document}
\draft
\title{How to generate families of spinors
}
\author{ N. Manko\v c Bor\v stnik}
\address{ Department of Physics, University of
Ljubljana, Jadranska 19,Ljubljana, 1111, \\
and Primorska Institute for Natural Sciences and Technology,\\
C. Mare\v zganskega upora 2, Koper 6000, Slovenia}
\author{ H. B. Nielsen}
\address{Department of Physics, Niels Bohr Institute,
Blegdamsvej 17,\\
Copenhagen, DK-2100
 }

\date{\today}

\maketitle

\begin{abstract}

Using a technique \cite{holgernorma2002} to construct a basis for spinors and ``families'' of
spinors in terms of Clifford algebra objects, we define other Clifford algebra objects, 
which transform the state of one ''family'' of spinors into the state of another ''family'' of spinors,
changing nothing but the ''family'' number.  
The proposed transformation  works - as does the technique - for all dimensions and any signature and
might open a path to understanding families of quarks and leptons\cite{norma92,norma93,normaixtapa2001,pikanorma2002}. 
\end{abstract}

\newpage
\section{Introduction}
\label{introduction}

We presented in a paper\cite{holgernorma2002} the technique to construct a spinor 
basis as products of nilpotents and projections  formed from the objects $\gamma^a$ for which we only 
need to know that they obey the Clifford algebra. Nilpotents and projections are odd and even objects 
of $\gamma^a$'s, respectively, and are chosen to be eigenstates of a Cartan subalgebra of the 
Lorentz group in the sense that the left multiplication of nilpotents and projectors by the Cartan 
subalgebra elements multiplies these objects by a number.  The technique can be used to 
construct a spinor basis for any dimension $d$
and any signature in a simple and transparent way. Equipped with graphic representation of basic states,  
the technique offers an elegant means of seeing all the quantum numbers of states with respect to the Lorentz 
group, as well as the transformation properties of states under Clifford algebra objects. 

Multiplying products of nilpotents and projectors from the left hand side  
by any of the Clifford algebra objects, we get a linear combination of these ``basic '' 
elements back: our basis spans a left ideal, and has $2^{d/2}$ elements for $d$ even and $2^{(d-1)/2}$
elements for $d$ odd. 

But there are $2^d$ products of nilpotents and projectors, all of them linearly independent. Mapping of 
ideals to spinor representations (treating
all  as Hilbert space) led accordingly  to $2^{d/2}$ replicas of the usual spinor 
representation for $d$ even and $d^{(d+1)/2}$ for $d$ odd. We called these replicas ''families'' of  
representations.

The proposed technique was initiated and developed by one of the authors of this paper, when  proposing an  
approach\cite{norma92,norma93,normaixtapa2001} in which all the internal degrees of freedom of either 
spinors or vectors can be described in the space of $d$-anticommuting (Grassmann) coordinates, if the 
dimension of ordinary space is also $d$.

In the approach of one of us, however, two kinds of $\gamma^a$ operators - two kinds of Clifford algebra
objects - were defined, both fulfilling the same Clifford algebra relations, while one kind anticommutes with 
the other kind. When one
of the two kinds of $\gamma^a$'s is used to generate nilpotents and projectors,  
the products of which define when operating in a vacuum state  basic vectors for the spinor representation of the
Lorentz group, it causes another kind of transition
among ''families'' of spinors, transforming one state of a ''family'' to a state with the same quantum numbers
with respect to the Lorentz group as the starting state but  belonging to another ''family''. 

We show in this short paper that there are also operators which cause transitions among ``families''\cite{pikanorma2002}
within the simple technique presented in \cite{holgernorma2002}. While left multiplication 
of nilpotents and projectors by Clifford algebra objects generates all the states of one spinor
representation (section \ref{technique}), {\em right multiplication causes transitions among ''families''}
(section \ref{families}).
(The reader should see also Chevalley's book \cite{chevalley}.) 

We demonstrate transitions among ``families'' for $d=4$ in subsection \ref{d4}.

In this paper, we  assume an arbitrary
signature of space time so that our metric tensor $\eta^{ab}$, with $a,b \in \{0,1,2,3,5,\cdots \d \}$ 
is diagonal with values $\eta^{aa} = \pm 1$, depending on the chosen signature ($+1$ for time-like coordinates
and $-1$ for space-like coordinates).

\section{Technique to generate spinor representations in terms of Clifford algebra objects}
\label{technique}

We shall briefly repeat the main points of the technique for generating spinor representations from 
Clifford algebra objects, following the
reference\cite{holgernorma2002}. We ask the reader to look for details and proofs in this reference.

We assume the objects $\gamma^a$, which fulfil the Clifford algebra
\begin{equation}
\{\gamma^a, \gamma^b \}_+ = I \;\;2\eta^{ab}, \quad {\rm for} \quad a,b \quad \in \{0,1,2,3,5,\cdots,d \},
\label{clif}
\end{equation}
for any $d$, even or odd.  $I$ is the unit element in the Clifford algebra, while
 $\{\gamma^a, \gamma^b \}_{\pm} = \gamma^a \gamma^b \pm \gamma^b \gamma^a $.

We assume  the ``Hermiticity'' property for $\gamma^a$'s 
\begin{eqnarray}
\gamma^{a\dagger} = \eta^{aa} \gamma^a,
\label{cliffher}
\end{eqnarray}
in order that 
$\gamma^a$ are compatible with (\ref{clif}) and formally unitary, i.e. ${\gamma^a}^{\dagger}\gamma^a=I$.

We also define the Clifford algebra objects 
\begin{equation}
S^{ab} = \frac{i}{4} [\gamma^a, \gamma^b ] := \frac{i}{4} (\gamma^a \gamma^b - \gamma^b \gamma^a)
\label{sab}
\end{equation}
which close the Lie algebra of the Lorentz group  
$ \{S^{ab},S^{cd}\}_- = i (\eta^{ad} S^{bc} + \eta^{bc} S^{ad} - \eta^{ac} S^{bd} - \eta^{bd} S^{ac})$.
One finds from Eq.(\ref{cliffher}) that $(S^{ab})^{\dagger} = \eta^{aa} \eta^{bb}S^{ab}$ and that $\{S^{ab}, S^{ac} \}_+
= \frac{1}{2} \eta^{aa} \eta^{bc}$.

Recognizing from Eq.(\ref{sab}) and the Lorentz algebra relation that two Clifford algebra objects 
$S^{ab}, S^{cd}$ with all indices different 
commute, we  select (out of infinitely many possibilities) the Cartan subalgebra of the algebra of the 
Lorentz group as follows 
\begin{eqnarray}
S^{0d}, S^{12}, S^{35}, \cdots, S^{d-2\; d-1}, \quad {\rm if } \quad d &=& 2n,
\nonumber\\
S^{12}, S^{35}, \cdots, S^{d-1 \;d}, \quad {\rm if } \quad d &=& 2n +1.
\label{choicecartan}
\end{eqnarray}

It is  useful  to define one of the Casimirs of the Lorentz group -  
the  handedness $\Gamma$ ($\{\Gamma, S^{ab}\}_- =0$). 
(For the definition of  $\Gamma$ for any spin in even-dimensional spaces, see 
references\cite{norma92,norma93,normaixtapa2001,bojannorma,%
holgernormadk}.) 
\begin{eqnarray}
\Gamma :&=&(i)^{d/2}\; \;\;\;\;\;\prod_a \quad (\sqrt{\eta^{aa}} \gamma^a), \quad {\rm if } \quad d = 2n, 
\nonumber\\
\Gamma :&=& (i)^{(d-1)/2}\; \prod_a \quad (\sqrt{\eta^{aa}} \gamma^a), \quad {\rm if } \quad d = 2n +1,
\label{hand}
\end{eqnarray}
for any integer $n$. We understand the product of $\gamma^a$'s in ascending order with respect to 
index $a$: $\gamma^0 \gamma^1\cdots \gamma^d$. 
It follows from Eq.(\ref{cliffher})
for any choice of the signature $\eta^{aa}$ that
$\Gamma^{\dagger}= \Gamma,\;
\Gamma^2 = I.$

We also find that for $d$ even, the handedness  anticommutes with the Clifford algebra objects 
$\gamma^a$ ($\{\gamma^a, \Gamma \}_+ = 0$) , while for $d$ odd it commutes with  
$\gamma^a$ ($\{\gamma^a, \Gamma \}_- = 0$). 

To make the technique simple, we introduce the graphic representation\cite{holgernorma2002}
as follows
\begin{eqnarray}
\stackrel{ab}{(k)}:&=& 
\frac{1}{2}(\gamma^a + \frac{\eta^{aa}}{ik} \gamma^b),\nonumber\\
\stackrel{ab}{[k]}:&=&
\frac{1}{2}(1+ \frac{i}{k} \gamma^a \gamma^b),\nonumber\\
\stackrel{+}{\circ}:&=& \frac{1}{2} (1+\Gamma),\nonumber\\
\stackrel{-}{\bullet}:&=& \frac{1}{2}(1-\Gamma),
\label{signature}
\end{eqnarray}
where $k^2 = \eta^{aa} \eta^{bb}$.
One can easily check by taking into account the Clifford algebra relation (Eq.\ref{clif}) and the
definition of $S^{ab}$ (Eq.\ref{sab})
that if one multiplies from the left hand side by $S^{ab}$ the Clifford algebra objects $\stackrel{ab}{(k)}$
and $\stackrel{ab}{[k]}$,
it follows that
\begin{eqnarray}
S^{ab}\stackrel{ab}{(k)}=\frac{1}{2}k \stackrel{ab}{(k)},\nonumber\\
S^{ab}\stackrel{ab}{[k]}=\frac{1}{2}k \stackrel{ab}{[k]},
\label{grapheigen}
\end{eqnarray}
which means that we get the same objects back multiplied by the constant $\frac{1}{2}k$. 
This also means that
$\stackrel{ab}{(k)}$ and $\stackrel{ab}{[k]}$ acting from the left hand side on anything (on a
vacuum state $|\psi_0\rangle$, for example ) are eigenvectors of $S^{ab}$.

We further find 
\begin{eqnarray}
\gamma^a \stackrel{ab}{(k)}&=&\eta^{aa}\stackrel{ab}{[-k]},\nonumber\\
\gamma^b \stackrel{ab}{(k)}&=& -ik \stackrel{ab}{[-k]}, \nonumber\\
\gamma^a \stackrel{ab}{[k]}&=& \stackrel{ab}{(-k)},\nonumber\\
\gamma^b \stackrel{ab}{[k]}&=& -ik \eta^{aa} \stackrel{ab}{(-k)}
\label{graphgammaaction}
\end{eqnarray}
It follows that
$
S^{ac}\stackrel{ab}{(k)}\stackrel{cd}{(k)} = -\frac{i}{2} \eta^{aa} \eta^{cc} 
\stackrel{ab}{[-k]}\stackrel{cd}{[-k]}, 
S^{ac}\stackrel{ab}{[k]}\stackrel{cd}{[k]} = \frac{i}{2}  
\stackrel{ab}{(-k)}\stackrel{cd}{(-k)}, 
S^{ac}\stackrel{ab}{(k)}\stackrel{cd}{[k]} = -\frac{i}{2} \eta^{aa}  
\stackrel{ab}{[-k]}\stackrel{cd}{(-k)}, 
S^{ac}\stackrel{ab}{[k]}\stackrel{cd}{(k)} = \frac{i}{2} \eta^{cc}  
\stackrel{ab}{(-k)}\stackrel{cd}{[-k]}.
$
It is useful to deduce the following relations
\begin{eqnarray}
\stackrel{ab}{(k)}^{\dagger}=\eta^{aa}\stackrel{ab}{(-k)},\quad
\stackrel{ab}{[k]}^{\dagger}= \stackrel{ab}{[k]},
\label{graphher}
\end{eqnarray}
and
\begin{eqnarray}
\stackrel{ab}{(k)}\stackrel{ab}{(k)}& =& 0, \quad \quad \stackrel{ab}{(k)}\stackrel{ab}{(-k)}
= \eta^{aa}  \stackrel{ab}{[k]}, \quad \stackrel{ab}{(-k)}\stackrel{ab}{(k)}=
\eta^{aa}   \stackrel{ab}{[-k]},\quad
\stackrel{ab}{(-k)} \stackrel{ab}{(-k)} = 0 \nonumber\\
\stackrel{ab}{[k]}\stackrel{ab}{[k]}& =& \stackrel{ab}{[k]}, \quad \quad
\stackrel{ab}{[k]}\stackrel{ab}{[-k]}= 0, \;\;\quad \quad  \quad \stackrel{ab}{[-k]}\stackrel{ab}{[k]}=0,
 \;\;\quad \quad \quad \quad \stackrel{ab}{[-k]}\stackrel{ab}{[-k]} = \stackrel{ab}{[-k]}
 \nonumber\\
\stackrel{ab}{(k)}\stackrel{ab}{[k]}& =& 0,\quad \quad \quad \stackrel{ab}{[k]}\stackrel{ab}{(k)}
=  \stackrel{ab}{(k)}, \quad \quad \quad \stackrel{ab}{(-k)}\stackrel{ab}{[k]}=
 \stackrel{ab}{(-k)},\quad \quad \quad 
\stackrel{ab}{(-k)}\stackrel{ab}{[-k]} = 0
\nonumber\\
\stackrel{ab}{(k)}\stackrel{ab}{[-k]}& =&  \stackrel{ab}{(k)},
\quad \quad \stackrel{ab}{[k]}\stackrel{ab}{(-k)} =0,  \quad \quad 
\quad \stackrel{ab}{[-k]}\stackrel{ab}{(k)}= 0, \quad \quad \quad \quad
\stackrel{ab}{[-k]}\stackrel{ab}{(-k)} = \stackrel{ab}{(-k)}.
\label{graphbinoms}
\end{eqnarray}
We recognize in  the first equation of the first row and the first equation of the second row
the demonstration of the nilpotent and the projector character of the Clifford algebra objects $\stackrel{ab}{(k)}$ and 
$\stackrel{ab}{[k]}$, respectively. 

{\em The reader should note that whenever the Clifford algebra objects apply from the left hand side,
they always transform } $\stackrel{ab}{(k)}$ {\em to} $\stackrel{ab}{[-k]}$, {\em never to} $\stackrel{ab}{[k]}$,
{\em and similarly } $\stackrel{ab}{[k]}$ {\em to} $\stackrel{ab}{(-k)}$, {\em never to} $\stackrel{ab}{(k)}$.

According to ref.\cite{holgernorma2002},  we define a vacuum state $|\psi_0>$ so that one finds
\begin{eqnarray}
< \;\stackrel{ab}{(k)}^{\dagger}
 \stackrel{ab}{(k)}\; > = 1, \nonumber\\
< \;\stackrel{ab}{[k]}^{\dagger}
 \stackrel{ab}{[k]}\; > = 1.
\label{graphherscal}
\end{eqnarray}

Taking the above equations into account it is easy to find a Weyl spinor irreducible representation
for $d$-dimensional space, with $d$ even or odd. (We advise the reader to see the reference\cite{holgernorma2002}.) 

For $d$ even, we simply set the starting state as a product of $d/2$, let us say, only nilpotents 
$\stackrel{ab}{(k)}$, one for each $S^{ab}$ of the Cartan subalgebra  elements (Eq.(\ref{choicecartan})),  applying it 
on an (unimportant) vacuum state\cite{holgernorma2002}. 
Then the generators $S^{ab}$, which do not belong 
to the Cartan subalgebra, applied to the starting state from the left hand side, 
 generate all the members of one
Weyl spinor.  
\begin{eqnarray}
\stackrel{0d}{(k_{0d})} \stackrel{12}{(k_{12})} \stackrel{35}{(k_{35})}\cdots \stackrel{d-1\;d-2}{(k_{d-1\;d-2})}
\psi_0 \nonumber\\
\stackrel{0d}{[-k_{0d}]} \stackrel{12}{[-k_{12}]} \stackrel{35}{(k_{35})}\cdots \stackrel{d-1\;d-2}{(k_{d-1\;d-2})}
\psi_0 \nonumber\\
\stackrel{0d}{[-k_{0d}]} \stackrel{12}{(k_{12})} \stackrel{35}{[-k_{35}]}\cdots \stackrel{d-1\;d-2}{(k_{d-1\;d-2})}
\psi_0 \nonumber\\
\vdots \nonumber\\
\stackrel{0d}{[-k_{0d}]} \stackrel{12}{(k_{12})} \stackrel{35}{(k_{35})}\cdots \stackrel{d-1\;d-2}{[-k_{d-1\;d-2}]}
\psi_0 \nonumber\\
\stackrel{od}{(k_{0d})} \stackrel{12}{[-k_{12}]} \stackrel{35}{[-k_{35}]}\cdots \stackrel{d-1\;d-2}{(k_{d-1\;d-2})}
\psi_0 \nonumber\\
\vdots 
\label{graphicd}
\end{eqnarray}
All the states have handedness $\Gamma $, since $\{ \Gamma, S^{ab}\}_- = 0$, which is easily calculated 
by multiplying from the left hand side the starting
state by $\Gamma$ of Eq.(\ref{hand}). 
States belonging to one multiplet  with respect to group $SO(q,d-q)$, that is to one
irreducible representation of spinors (one Weyl spinor), can have any phase. We chose 
the simplest one, setting all  phases equal to one.

The above graphic representation demonstrated that for $d$ even 
all the states of one irreducible Weyl representation of a definite handedness follow from the starting state, 
which is, for example, a product of nilpotents $\stackrel{ab}{(k)}$, by transforming all possible pairs
of $\stackrel{ab}{(k)} \stackrel{mn}{(k)}$ into $\stackrel{ab}{[-k]} \stackrel{mn}{[-k]}$.
There are $S^{am}, S^{an}, S^{bm}, S^{bn}$, which do this.
The procedure gives $2^{(d/2-1)}$ states. A Clifford algebra object $\gamma^a$ applied from the left hand side
transforms  a 
Weyl spinor of one handedness into a Weyl spinor of the opposite handedness. Both Weyl spinors form a Dirac 
spinor. We call such a set of
states a ''family''. 

For $d$ odd a Weyl spinor also has in addition to a product of $(d-1)/2$ nilpotents or projectors either the
factor $\stackrel{+}{\circ}:= \frac{1}{2} (1+\Gamma)$ or the factor
$\stackrel{-}{\bullet}:= \frac{1}{2}(1-\Gamma)$. (See the ref.\cite{holgernorma2002}.) 
As in the case of $d$ even, all the states of one irreducible 
Weyl representation of a definite handedness follow from a starting state, 
which is, for example, a product of $\frac{1}{2}(1 + \Gamma)$ and $(d-1)/2$ nilpotents $\stackrel{ab}{(k)}$, by 
transforming all possible pairs
of $\stackrel{ab}{(k)} \stackrel{mn}{(k)}$ into $\stackrel{ab}{[-k]} \stackrel{mn}{[-k]}$.
But $\gamma^a$'s applied from the left hand side do not change the handedness of the Weyl spinor, 
since $\{ \Gamma,
\gamma^a \}_- =0$ for $d$ odd\cite{holgernorma2002}.
A Dirac and a Weyl spinor are for $d$ odd identical and a ''family'' 
has accordingly $2^{(d-1)/2}$ members of basic states of a definite handedness.

We shall speak about left handedness when $\Gamma= -1$ and right
handedness when $\Gamma =1$ for either $d$ even or odd. 

{\em When the whole Clifford algebra is considered as states in a Hilbert space, then we get ''families''}.

\section{``Families'' }
\label{families}

When all $2^d$ states are considered as a Hilbert space, we recognize that for $d$ even 
there are $2^{d/2}$ ''families'' and for $d$ odd $2^{(d+1)/2}$ ''families'' of spinors.

We prove  in this section (see also \cite{pikanorma2002}) that there exists an operation 
which transforms the state of one ''family''
into the state of another ''family'', leaving all the  properties with respect to the Lorentz group unchanged.

We saw in the last section (\ref{technique}) that any Clifford algebra object when
multiplying from the left hand side products of nilpotents and projectors 
(operating on a vacuum state) - a state of a Dirac spinor - transforms this state into a 
superposition of states of the same Dirac spinor. We  refer to a Dirac spinor as a ''family''. Since there are $2^d$ 
linearly independent states, one finds for $d$ even  $2^{d/2}$ ''families''  and for $d$ odd $2^{(d+1)/2}$ 
''families'' of Dirac spinors.
''Families'' form left ideals with respect to the multiplication with the Clifford algebra objects.

{\em The question then arises: Which operation  transforms the state of one ''family'' into the state of
another ''family''?}

{\em Statement 1}: {\em Right multiplication with the Clifford algebra objects transforms the state of one
''family'' into the state of another ''family''}.

 {\em Proof:} The Clifford algebra object $\stackrel{ab}{(k)}$ transforms, when applied from the left
hand side 
by either $\gamma^a$ (or by $\gamma^b$), into $\stackrel{ab}{[-k]}$ (Eq.(\ref{graphgammaaction})). 
One finds this by simply multiplying 
$\stackrel{ab}{(k)}$ from the left hand side by one of these two Clifford algebra objects and 
taking into account Eq.(\ref{clif})
\begin{eqnarray}
\gamma^a \stackrel{ab}{(k)} = \gamma^a \; \frac{1}{2}(\gamma^a + \frac{\eta^{aa}}{ik}\gamma^b)= \eta^{aa}
\frac{1}{2}(1  + \frac{i}{-k}\gamma^a \gamma^b) = \eta^{aa} \stackrel{ab}{[-k]}.
\label{leftcheck}
\end{eqnarray}
(And similarly we get $\gamma^b \stackrel{ab}{(k)}= -ik \stackrel{ab}{[-k]}$. The product of $\gamma^a \gamma^b$ 
transforms $\stackrel{ab}{(k)}$ into itself, multiplying it by  $-ik$, since $\stackrel{ab}{(k)}$ was chosen to be 
the ''eigen vector'' of the Cartan subalgebra of the Lorentz algebra in the sense.)

Let us now multiply the same object $\stackrel{ab}{(k)}$ by  $\gamma^a$  {\em from
the right hand side}. It follows
\begin{eqnarray}
\stackrel{ab}{(k)}\gamma^a =  \frac{1}{2}(\gamma^a + \frac{\eta^{aa}}{ik}\gamma^b)\; \gamma^a = \eta^{aa}
\frac{1}{2}(1  + \frac{i}{k}\gamma^a \gamma^b) = \eta^{aa} \stackrel{ab}{[k]}.
\label{rightcheck}
\end{eqnarray}
We saw in Eq.(\ref{leftcheck}) that multiplication from the left hand side by any Clifford algebra object
transforms $\stackrel{ab}{(k)}$ into $\stackrel{ab}{[-k]}$, never to $\stackrel{ab}{[k]}$.
This means that $\stackrel{ab}{[k]}$ is going to be a building block of a different ''family'' than
$\stackrel{ab}{[-k]}$. 

{\em Theorem 1:} {\em The two operations - left and right multiplication by $\gamma^a$ - commute}. 

{\em Proof:} To see this we need to show that the two objects $\stackrel{ab}{(k)}$ 
and $\stackrel{ab}{[k]}$, the second obtained from the first by right multiplication,
have all the properties with respect to the Lorentz group (application of the Lorentz algebra 
objects concerns the left multiplication) equal and that they
differ only in the ''family'' name. Left multiplication by $\gamma^a $ of the object
$\stackrel{ab}{(k)}$ leads, as we know (Eq(\ref{grapheigen})), to $\stackrel{ab}{[-k]}$, 
whose $S^{ab} \stackrel{ab}{[-k]}= -\frac{k}{2} \stackrel{ab}{[-k]}$.

To check the properties of the two Clifford algebra objects $\stackrel{ab}{(k)}$ and 
$\stackrel{ab}{[k]}$ with respect to the Lorentz group, we have to multiply each of the
two Clifford algebra objects  from 
the left hand side by $S^{ab}$, which is the Cartan subalgebra element. 
According to Eq.(\ref{grapheigen})
we find
$S^{ab}\stackrel{ab}{(k_{ab})} = \frac{i}{2} \frac{1}{2}(\gamma^a \gamma^b \gamma^a + \frac{\eta^{aa}}{ik}
\gamma^a \gamma^b \gamma^b) = \frac{k}{2}  \stackrel{ab}{(k)},\;
S^{ab}\stackrel{ab}{[k]} = \frac{k}{2} k \stackrel{ab}{[k]}.
$

Both objects, $\stackrel{ab}{(k)}$ and $\stackrel{ab}{[k]}$, have the same eigenvalue for the Cartan 
subalgebra element $S^{ab}$, namely $\frac{1}{2} k $. Since right multiplication of the object
$\stackrel{ab}{(k)}$ does not change the properties of the object with respect to the Lorentz group 
(the properties of which are determined by left multiplication) the two operations - 
left and right multiplication
with $\gamma^a$'s, both fulfilling the Clifford algebra relation - must commute 
and the proof is completed.

Since $\gamma^a$'s are odd Clifford algebra objects, we would like to see the two operations - left and 
right multiplication with $\gamma^a$'s - to anticommute rather then commute. With appropriate choice of a phase
we can make them commute.

{\em We define}\cite{pikanorma2002} {\em the Clifford algebra objects}
$\tilde{\gamma}^a$'s {\em as operations which operate formally from the left hand side} (as $\gamma^a$'s do)
{\em on objects } $\stackrel{ab}{(k)}$ {\em and} $\stackrel{ab}{[k]}$, {\em transforming objects to}
$\stackrel{ab}{[k]}$ {\em and} $\stackrel{ab}{(k)}$, respectively, {\em as} $\gamma^a$'s {\em would 
if applied from the right hand side, up to a phase i}
\begin{eqnarray}
\tilde{\gamma^a} \stackrel{ab}{(k)}: &=& -i\stackrel{ab}{(k)} \gamma^a = - i\eta^{aa}\stackrel{ab}{[k]},\noindent\\
\tilde{\gamma^b} \stackrel{ab}{(k)}: &=& -i\stackrel{ab}{(k)} \gamma^b = - k
\stackrel{ab}{[k]}.
\label{gammatilde}
\end{eqnarray}
One accordingly finds
\begin{eqnarray}
\tilde{\gamma^a} \stackrel{ab}{[k]}: &=& \;\; i \stackrel{ab}{[k]} \gamma^a = \;\;i\stackrel{ab}{(k)},\noindent\\
\tilde{\gamma^b} \stackrel{ab}{[k]}: &=& \;\; i \stackrel{ab}{[k]} \gamma^b = -k \eta^{aa} \stackrel{ab}{(k)}.
\label{gammatilde1}
\end{eqnarray}
We generalize the above definition of $\tilde{\gamma^a}$ to any Clifford algebra object as follows
\begin{eqnarray}
\tilde{\gamma^a} A = i (-)^{(A)} A \gamma^a,
\label{gammatildeA}
\end{eqnarray}
with $(-)^{(A)} = -1$,  if $A$ is an odd Clifford algebra object and 
$(-)^{(A)} = 1$,  if $A$  is an even Clifford algebra object.

We can prove that $\tilde{\gamma^a}$ obey the same Clifford algebra relation as $\gamma^a$.
\begin{eqnarray}
(\tilde{\gamma^a} \tilde{\gamma^b} + \tilde{\gamma^b} \tilde{\gamma^a}) A = -i i ((-)^{(A)})^2 A (
\gamma^a \gamma^b + \gamma^b \gamma^a) = 2\eta^{ab} A
\label{gammatildeACL}
\end{eqnarray}
and that $\tilde{\gamma^a}$ and $\gamma^a$ anticommute
\begin{eqnarray}
(\tilde{\gamma^a} \gamma^b + \gamma^b \tilde{\gamma^a}) A =  i(-)^{(A)}( -\gamma^b A \gamma^a 
 + \gamma^b A \gamma^a)  = 0.
\label{gammatildeACL}
\end{eqnarray}

 From {\em Theorem 1} and the above calculations, we may write
\begin{eqnarray}
\{ \tilde{\gamma^a}, \gamma^b \}_+ &=& 0, \quad {\rm while} \quad
\{\tilde{\gamma^a}, \tilde{\gamma^b} \}_+ = 2 \eta^{ab}.
\label{gammatildegamma}
\end{eqnarray}
If we define
\begin{eqnarray}
\tilde{S}^{ab} = \frac{i}{4}\; [\tilde{\gamma}^a,\tilde{\gamma}^b] = \frac{1}{4} (\tilde{\gamma}^a
\tilde{\gamma}^b - \tilde{\gamma}^b\tilde{\gamma}^a),
\label{tildesab}
\end{eqnarray}
it follows
\begin{eqnarray}
\tilde{S}^{ab} A = A \frac{1}{4} (\gamma^b \gamma^a - \gamma^a \gamma^b),
\label{tildesab1}
\end{eqnarray}
manifesting accordingly that $\tilde{S}^{ab}$ fulfil the Lorentz algebra relation as $S^{ab}$ do. Taking into account 
Eq.(\ref{gammatildeA}), we further find
\begin{eqnarray}
\{\tilde{S}^{ab}, S^{ab}\}_- =0,\quad
\{\tilde{S}^{ab}, \gamma^c \}_-=0,\quad
\{S^{ab}, \tilde{\gamma}^c \}_-=0.
\label{sabtildesab}
\end{eqnarray}
One also finds 
\begin{eqnarray}
\{\tilde{S}^{ab}, \Gamma \}_- =0,\quad \{ \tilde{\gamma}^a, \Gamma \}_- = 0, \quad {\rm for \;\; d\;\; even,}\nonumber\\
\{\tilde{S}^{ab}, \Gamma \}_- =0,\quad \{ \tilde{\gamma}^a, \Gamma \}_+ = 0, \quad {\rm for \;\; d\;\; odd,}
\label{sabtildeGAMMA}
\end{eqnarray}
which means that in $d$ even transforming one ''family'' into another with either $\tilde{S}^{ab}$ 
or $\tilde{\gamma}^a$ leaves handedness $\Gamma$ unchanged, while the transformation to another 
''family'' in $d$ odd with $\tilde{\gamma}^a$ changes the handedness of states, namely the factor
 $\frac{1}{2}(1\pm \Gamma)$ changes to  $\frac{1}{2}(1\mp \Gamma)$ in accordance with what we know from before:
 In spaces with odd $d$  changing the handedness means changing the ''family''.

We advise the reader also to read \cite{norma93,normaixtapa2001} where the two kinds of 
Clifford algebra objects follow as two different superpositions of a Grassmann coordinate and its 
conjugate momentum.

We present  for $\tilde{S}^{ab}$ some useful relations
\begin{eqnarray}
\tilde{S}^{ab} \stackrel{ab}{(k)}&=& \;\;\frac{k}{2}\stackrel{ab}{(k)},\nonumber\\
\tilde{S}^{ab} \stackrel{ab}{[k]}&=& -\frac{k}{2} \stackrel{ab}{[k]}, \nonumber\\
\tilde{S}^{ac}\stackrel{ab}{(k)}\stackrel{cd}{(k)}& = &\frac{i}{2} \eta^{aa} \eta^{cc} 
\stackrel{ab}{[k]}\stackrel{cd}{[k]}, \nonumber\\
\tilde{S}^{ac}\stackrel{ab}{[k]}\stackrel{cd}{[k]}& = & -\frac{i}{2}  
\stackrel{ab}{(k)}\stackrel{cd}{(k)}, \nonumber\\
\tilde{S}^{ac}\stackrel{ab}{(k)}\stackrel{cd}{[k]}& = & -\frac{i}{2} \eta^{aa}  
\stackrel{ab}{[k]}\stackrel{cd}{(k)}, \nonumber\\
\tilde{S}^{ac}\stackrel{ab}{[k]}\stackrel{cd}{(k)}& = &\frac{i}{2} \eta^{cc}  
\stackrel{ab}{(k)}\stackrel{cd}{[k]}.
\label{tildesac}
\end{eqnarray}

According to {\em Statement 1} we transform the state of one ''family'' to the state of another ''family'' by 
the application of
$\tilde{\gamma}^a$ or $\tilde{S}^{ac}$ (formally from the left hand side) on a state of the first
''family'' for a chosen $a$ or $a,c$. To transform all the states of one ''family'' into states
of another ''family'', we apply  $\tilde{\gamma}^a$ or $\tilde{S}^{ac}$ to each state of the starting ''family''.
It is, of course, sufficient to apply $\tilde{\gamma}^a$ or $\tilde{S}^{ac}$ 
to only one state of a ''family'' and then use generators of the Lorentz group ($S^{ab}$), and for $d$ even 
also $\gamma^a$'s, 
to generate all the states of one Dirac spinor.

One must notice that nilpotents $\stackrel{ab}{(k)}$ and projectors $\stackrel{ab}{[k]}$ are eigenvectors
not only of the Cartan subalgebra $S^{ab}$ but also of $\tilde{S}^{ab}$. Accordingly only $\tilde{S}^{ac}$, which
do not carry the Cartan subalgebra indices, cause the transition from one ''family'' to another ''family''.

The starting state of Eq.(\ref{graphicd}) can change, for example, to
\begin{eqnarray}
\stackrel{0d}{[k_{0d}]} \stackrel{12}{[k_{12}]} \stackrel{35}{(k_{35})}\cdots \stackrel{d-1\;d-2}{(k_{d-1\;d-2})},
\label{tildegraphicd}
\end{eqnarray}
if $\tilde{S}^{01}$ was chosen to transform the Weyl spinor of Eq.(\ref{graphicd}) to the Weyl spinor of another
''family''.

In the next section \ref{d4} we demonstrate the appearance of ''families for $d=4$ with the Minkowski
signature.

\subsection{''Families'' for $d=4$}
\label{d4}

There are two  $(d/2=2)$ operators of the Cartan subalgebra of the Lorentz algebra (which is 
closed by the operators 
$S^{01}, S^{02}, S^{03}, S^{12}, S^{13}, S^{23}$), for which we made a choice, according to 
Eq.(\ref{choicecartan})
of $ S^{03}$ and $S^{12}$. Following Eq.(\ref{hand}) 
we find $\Gamma = 
i \gamma^0 \gamma^1 \gamma^2 \gamma^3. $ There are $2^4$ , that is sixteen basic states,
all of them  being  ''eigenstates'' of $S^{12}$ and $S^{03}$ in the sense that Eq.(\ref{grapheigen})
\begin{eqnarray}
\stackrel{03}
{(\pm i)} \stackrel{12}{(\pm)}&=&(\frac{1}{2})^2 \;(\gamma^0 \mp \gamma^3) \;(\gamma^1 \pm i \gamma^2), \quad 
\stackrel{03}
{(\pm i)} \stackrel{12}{[\pm]}\;=\;(\frac{1}{2})^2 \;
(\gamma^0 \mp \gamma^3) (1 \pm i \gamma^1 \gamma^2),
\nonumber\\
\stackrel{03}
{[\pm i]} \stackrel{12}{(\pm)}&=&(\frac{1}{2})^2 (1 \pm \gamma^0 \gamma^3)\; (\gamma^1 \pm i \gamma^2), \quad 
\stackrel{03}
{[\pm i]} \stackrel{12}{[\pm]}\;=\;(\frac{1}{2})^2
(1 \pm \gamma^0 \gamma^3) (1 \pm i \gamma^1 \gamma^2),
\label{basisfour}
\end{eqnarray}
with the ''eigenvalues'' of the Cartan operator $S^{12}$ equal to
 $\pm 1/2$ for the $\pm$ sign in the second factor of the graphical presentation and the eigenvalues of the Cartan operator 
$S^{03}$ equal to
$\pm i/2$ for the $\pm i$ sign in the first factor of the graphical representation. All sixteen basic states 
are orthonormal\cite{holgernorma2002}.

We arrange these sixteen states into four ``families'' by first choosing the starting state of
the first ''family'' as a product of two nilpotents $\stackrel{03}
{(+i)} \stackrel{12}{(+)} $.  Then we use $S^{01}$, for example, to find the second state 
$\stackrel{03}
{[-i]} \stackrel{12}{[-]} $ of one Weyl spinor. $\gamma^0$ generates from the first state of the first Weyl spinor
the first state of the second Weyl spinor, namely $\stackrel{03}{[-i]} \stackrel{12}{(+)} $ and $S^{01}$ then the
second state of the second spinor $\stackrel{03}{(+i)} \stackrel{12}{[-]} $. We set all the phases of 
the states equal to one.

We transform this first ''family'' into the second ''family'' by applying $\tilde{S}^{01}$ 
(or $\tilde{S}^{02}$, or $\tilde{S}^{31}$, or $\tilde{S}^{32}$) to each of the
states of the first ''family''. By applying $\tilde{\gamma}^a$ to all the states of the first ''family''
we get the third ''family''. We also can generate the third ''family'' from the second by applying 
$\tilde{\gamma}^1$ (or $\tilde{\gamma}^{2}$ ). The fourth ''family'' can be reached from the first one by the application of
$\tilde{\gamma}^1$ (or $\tilde{\gamma}^2$), but it can also be reached from any other with the appropriate 
choice of operations. For all the ''families'' the simplest choice of the relative phases, namely the phase $1$ 
is made.

Each ``family'' includes two Weyl spinors, one left- and one right-handed. 
These four ``families'' are  presented in Table I.

\begin{center}
\begin{tabular}{|r|r||c|c|c|c||r|r|r||}
\hline
a&i&$(^a\psi_i)_1$&$(^a\psi_i)_2$&$(^a\psi_i)_3$&$(^a\psi_i)_4$&$ S^{12}$&$ S^{03}$&$
\Gamma$\\
\hline\hline
1&1&$\quad \stackrel{03}{(+i)} \stackrel{12}{(+)} $&$\quad \stackrel{03}{[+i]} \stackrel{12}{[+]} $&
$\quad \stackrel{03}{[+i]} \stackrel{12}{(+)}$&$\quad \stackrel{03}{(+i)} \stackrel{12}{[+]} $& 
$\frac{1}{2}$& $\frac{i}{2}$& -1\\
\hline 
1&2&$\quad \stackrel{03}{[-i]} \stackrel{12}{[-]} $&$\quad \stackrel{03}{(-i)} \stackrel{12}{(-)} $&
$\quad \stackrel{03}{(-i)} \stackrel{12}{[-]}$&$\quad \stackrel{03}{[-i]} \stackrel{12}{(-)} $&
 $-\frac{1}{2}$& $-\frac{i}{2}$& -1\\
\hline\hline
2&1&$\quad \stackrel{03}{[-i]} \stackrel{12}{(+)} $&$\quad \stackrel{03}{(-i)} \stackrel{12}{[+]} $&
$\quad \stackrel{03}{(-i)} \stackrel{12}{(+)}$&$\quad \stackrel{03}{[-i]} \stackrel{12}{[+]} $
& $\frac{1}{2}$& $-\frac{i}{2}$& 1\\
\hline 
2&2&$\quad \stackrel{03}{(+i)} \stackrel{12}{[-]} $&$\quad \stackrel{03}{[+i]} \stackrel{12}{(-)} $&
$\quad \stackrel{03}{[+i]} \stackrel{12}{[-]}$&$\quad \stackrel{03}{(+i)} \stackrel{12}{(-)} $
& $-\frac{1}{2}$& $\frac{i}{2}$& 1\\
\hline\hline
\end{tabular}
\end{center}
Table I.-Four ``families'' of the two Weyl spinors of the Lorentz group $SO(1,3)$.
Basic vectors are eigenvectors of the two  operators of the Cartan subalgebra
$S^{12}$ and $S^{03}$. The eigenvalues of the operator of handedness $\Gamma $
are also presented. All the basic states are orthonormalized as discussed in ref.\cite{holgernorma2002}.
The simplest choice of  relative phases is used - 
all phases are assumed to be  equal to $+1$. 

We see in Table I that either $\tilde{\gamma}^a, \; a=0,1,2,3$, or $\tilde{S}^{01}, \tilde{S}^{02},
\tilde{S}^{31},\tilde{S}^{32},$ when applied, change the ''family'' but do not change either 
the handedness or the
''eigenvalues'' of the Cartan subalgebra of the Lorentz algebra.

Any of the four ``families'' can be used to represent the solution of the Dirac  equation
for a massive spinor, while massless spinors are either left- or right-handed, so that
only half of the space of the massive case is needed to find the solution. The phases chosen for basic states make
the matrix representation of $\gamma^a$'s and $S^{ab}$ equal to the usual ones.

\section{Conclusion}
\label{conclusion}

In \cite{holgernorma2002} we constructed the basis for a left ideal out of products of nilpotents
and projectors and identified the basis with 
the spinor space. There are $2^d$ nilpotents and projectors. Mapping all the ideals to spinor representations, 
that is treating all as a Hilbert space, leads to ''families'' of spinors. An irreducible representation of  a
spinor (a Dirac spinor, which is a Weyl bi-spinor with $2^{d/2}$ members for $d$ even
and a Weyl spinor with $2^{(d-1)/2}$ members for $d$ odd) depends on a selection of a ``starting'' Clifford object. 
We have for $d$ even $2^{d/2}$ different starting states and for $d$ odd $2^{(d+1)/2}$ different 
starting states, which  can all be made orthogonal 
with the appropriate choice of an (unimportant)
vacuum state. These are the different bases (that is different spinor spaces, which have 
all the same properties with respect to the Lorentz group), which we call   ''families'' of spinors.

We have shown in 
this paper that while left multiplication with any Clifford algebra object makes from a ''starting'' state 
a superposition of basic states of one Dirac spinor, {\em right multiplication} of a ''starting state''
with any Clifford algebra object makes a superposition of states belonging to different ''families'' but having the 
same properties with respect 
to the Lorentz group (defined by left multiplication of   Clifford algebra objects)  as the starting state.
One comes  accordingly from one ''family'' to another ''family'' (up to an overall factor) by multiplying 
the ''starting'' state from the right hand side by $S^{ab}$'s for $d$ odd and by $S^{ab}$'s and $\gamma^a$'s 
for $d$ even, if $S^{ab}$ does not belong to the Cartan subalgebra of the Lorentz group.

We have defined in this paper $\tilde{\gamma}^{a}$'s and $\tilde{S}^{ab}$ with the properties that both
transform nilpotents and projectors, when multiplying them from the left hand side as
$\gamma^{a}$'s and $S^{ab}$ would if multiplying nilpotents and projectors from the right hand side. Since neither
$\tilde{\gamma}^{a}$'s nor $\tilde{S}^{ab}$  change properties of nilpotents and projectors (and
consequently of a state) with respect to the Lorentz group, and since both - $\gamma^a$'s and 
$\tilde{\gamma}^a$'s - are Clifford algebra objects, 
 it follows (with the appropriate choice of phases) that $\{ \tilde{\gamma}^{a},\gamma^b \}_+
=0$ and  $\{ \tilde{\gamma}^{a},\tilde{\gamma}^b \}_+ =
\eta^{ab}$. Consequently, $\tilde{S}^{ab}$ fulfil the same Lorentz algebra relation as $S^{ab}$. This can 
be understood since one could construct the spinor basis for a right ideal instead of a left ideal and then use left
multiplication to generate ``families''. (To understand this better, see 
\cite{norma93,normaixtapa2001,holgernormadk,pikanorma2002}.)

We have also demonstrated the application of $\tilde{\gamma}^a$'s and $\tilde{S}^{ab}$ on a basis with the help
of the graphic technique introduced in ref.\cite{holgernorma2002}. We demonstrated the procedure 
for generating ''families'' in the case of $d =1 +3$, that is for $d=4$ and one time coordinate, where the number
of ''families'' is four.

In conclusion, we must ask ourselves whether the proposed generation of  ''families'' 
can be used to describe the ``families'' of 
quarks and leptons? We believe that we have the right way to do this. According 
to refs.\cite{normaixtapa2001,pikanorma2002}, one can generate ``families'' of quarks and leptons dynamically if in the
covariant derivative $\tilde{S}^{ab} $ appear as charges, accompanied by gauge fields like 
$\tilde{\omega}^{ab}{}_{c}$, so that 
\begin{eqnarray}
p^a{}_o = p^a - \frac{1}{2} \tilde{S}^{ab} \tilde{\omega}_{abc}- \tau^{Ai} A^{Ai a},
\label{covariantp}
\end{eqnarray}
with $\tau^{Ai}$ determining the known charges ($U(1), SU(2), SU(3) $) and $A^{Ai a}$ the corresponding gauge fields.
This possibility has been discussed  in ref.\cite{pikanorma2002}.

In the presented formalism we worked with the Clifford algebra objects only, using them to express not only 
the generators of the Lorentz algebra but also the operators transforming one ``family'' into another ``family'' and  
even the spinor basis for one and several ``families''. (It is for this reason that we get the 
``families´´ as explained.) 
One would accordingly ask oneself if there is 
possibly any physical reason that the Clifford algebra degrees of freedom are ``more fundamental'' than the spinors.

This is exactly what was taken as the starting point in the works by one of 
us\cite{norma92,norma93,normaixtapa2001,pikanorma2002} (and which lead to the formalism, presented in this paper)
but it is also what is really the case in the Kogut and Susskind lattice fermion formalism\cite{kogutsusskind}. 
In the latter case the fermion degrees of freedom are interpreted as sitting really on
the sites links plaquettes and cubes and hypecubes of a lattice with the double lattice constant. That 
corresponds to having fields of all the possible antisymmetrised tensor characters, which is just what the
Clifford algebra elements have. One therefore has also in the Kogut-Susskind fermions a way of having the 
Clifford algebra naturally more fundamental than the spinors. Not surprisingly we get also in this case
the expected number $4$ for families in this lattice model. 
Also the paper of both of us\cite{dirackahler} already introduced ``families'' of spinors with the help 
of the Clifford algebra objects, but in a slightly different way - the spinor states have two indices.

\section{Acknowledgement }

This work was supported by the Ministry of Education, 
Science and Sport of Slovenia  as well as by funds CHRX -
CT - 94 - 0621, INTAS 93 - 3316, INTAS - RFBR 95
- 0567, SCI-0430-C (TSTS) of  Denmark.

\end{document}